\documentclass[journal abbreviation, manuscript]{copernicus}

\usepackage{graphicx}
\usepackage{hyperref}
\usepackage{booktabs}

\begin{document}

\nolinenumbers

\title{Tracking Summer Greenland Blocking: the Upstream Pathway Shapes Historical Extremes and Future Change}


\Author[1,2][michele.filippucci@unitn.it]{Michele}{Filippucci} 
\Author[3]{Jacob}{Maddison}
\Author[1,4][simona.bordoni@unitn.it]{Simona}{Bordoni}

\affil[1]{Department of Civil, Environmental and Mechanical Engineering, University of Trento, Trento, Italy}
\affil[2]{Istituto Universitario di Studi Superiori (IUSS) di Pavia, Pavia, Italy}
\affil[3]{Department of Mathematics and Statistics, University of Exeter, Exeter, UK}
\affil[4]{Center Agriculture Food Environment (C3A), University of Trento, San Michele all’Adige, Italy}



\runningtitle{TEXT}

\runningauthor{TEXT}

\received{}
\pubdiscuss{} 
\revised{}
\accepted{}
\published{}


\firstpage{1}

\maketitle

\begin{abstract}
The representation and future evolution of summer Greenland atmospheric blocking in climate models is here investigated from a Lagrangian perspective using a novel Python package \textit{blocktrack}. By applying the blocktrack algorithm to ERA5 reanalysis and a CMIP6 model ensemble, we identify and track blocking events over Greenland, and obtain their trajectories, intensities, duration and wave-breaking patterns. Greenland blocking (GB) events in ERA5 are then classified into two types  based on their wave-breaking characteristics. These correspond to the previously identified upstream (anticyclonic wave breaking) and retrograding (cyclonic wave breaking) GBs. Upstream blocks, which originate in Northern Canada, exhibit stronger moisture transport before and during blocking onset and higher temperature anomalies than retrograding blocks, which follow an east-to-west trajectory and originate in the North Atlantic. Our analyses show how the recent observed increase in GB frequency, particularly in 2012, is primarily driven by upstream blocks.
CMIP6 models generally fail to capture the observed increase and underestimate GB variability, especially for the upstream component. Projections under the SSP3-7.0 scenario show a decline in retrograding blocks but a possible increase in upstream blocks, depending on the detection index used. We discuss possible drivers of these changes, which include jet stream shifts, increased frequency of high-moisture transport events from low to high latitudes, surface temperature increases due to Atlantic Multidecadal Variability and Arctic Amplification. By analyzing block trajectories, this study demonstrates how Lagrangian diagnostics can provide novel insights into the dynamics of blocking events over Greenland.
\end{abstract}


\section{Introduction}

The climate of Greenland and the dynamics of its ice sheet are of critical importance in the context of global warming due to their significant contribution to sea level rise and their role in the Atlantic Meridional Overturning Circulation (AMOC). If the Greenland ice sheet were to become unstable and initiate irreversible melting, the ocean level would rise by up to 7 meters over the next millennium \citep{gregory2004threatened}, with severe impacts on coastal regions worldwide. Freshwater input into the North Atlantic ocean would also alter high-latitude salinity, potentially disrupting the AMOC and the associated heat transport from low to high latitudes \citep{driesschaert2007modeling,buckley2016observations}. Such disruption would have major consequences on Northern Hemisphere (NH) climate, particularly northern Europe \citep{bellomo2021future,vacca2025role}. Between 2010 and 2020, Greenland has experienced exceptional episodes of surface melt and runoff, culminating in record-breaking melt years in 2012 and 2019 \citep{tedesco2020unprecedented}. These extreme melt events occur predominantly during the summer months, when solar radiation is highest and surface energy input drives peak ice sheet melt. Understanding the atmospheric drivers of these anomalies is essential for improving projections of ice sheet mass balance and circulation responses to global warming. 

Extreme temperatures and record ice melt in Greenland are strongly influenced by Greenland Blocking (GB) \citep{hanna2013influence}. In general terms, atmospheric blocking refers to quasi stationary high-pressure systems that disrupt the typical zonal flow of the mid-latitude jet stream, especially in the NH \citep{charney1979multiple,hoskins1985use}. GB, specifically, is characterized by persistent high-pressure anomalies over or near Greenland \citep{vautard1990multiple} and plays a key role in modulating both the local weather of Greenland and the larger-scale circulation patterns across the North Atlantic region. In fact, GB variability is strongly anticorrelated with the North Atlantic Oscillation (NAO), the most prominent pattern of climate variability in the extratropical North Atlantic region \citep{benedict2004synoptic,woollings2008new,davini2012coupling}.

The frequency of GB has been observed to increase during the first two decades of the 21st century, with very large anomalies around 2012 and 2019 coinciding with major melt years \citep{tedesco2013evidence,hanna2016greenland,mcleod2016linking} and a decrease after 2020 \citep{preece2023summer,maddison2024missing}. This suggests a link between the occurrence of blocking and anomalous heating and increased melting. Suggested mechanisms behind the observed increase include hemispheric-scale teleconnection patterns manifesting as a stationary Rossby wave with a zonal wave number-3 pattern \citep{teng2022warming}, reductions in North American snow cover linked to Arctic Amplification \citep{preece2023summer}, and sea surface temperature (SST) anomalies, particularly in association with the Atlantic Multidecadal Oscillation (AMO) \citep{beckmann2025summer}. Ultimately, these different mechanisms are believed to alter the climatological geopotential height west of Greenland, creating a stationary wave that promotes blocking over the region. Previous studies have also explored the role of precursor cyclones in advecting warm air over Greenland and amplifying blocking anticyclones leading to extreme blocking events \citep{mcleod2015assessing}. Additionally, moisture advection into the Greenland region has been identified as an amplifier of blocking-induced ice melt \citep{liu2015extreme,barrett2020extreme,mattingly2023increasing}. Similarly, the co-occurrence of atmospheric rivers and foehn events over northeast Greenland has been linked to extreme melt episodes by \citet{mattingly2023increasing}. Moreover, recent work \citep{luu2024greenland,maddison2024missing} has shown that neither the CMIP6 large-ensemble simulations nor HIGHRESMIP models are able to reproduce the observed recent increase in GB frequency, suggesting that current models are missing important factors governing GB variability. Finally, \citet{wachowicz2021historical} have shown that the observed GB increase and interannual variability depends on the blocking index adopted and that statistically significant increases emerge only when a 5-year running mean is applied to the frequency time series. Despite extensive research on this topic, we still lack a definitive explanation for both the recent increase in GB events and the inability of climate models to capture it.

Some of the difficulties encountered in interpreting the recent summer blocking increase are a consequence of the existing gaps in our understanding of GB. To address this issue, \citet{hauser2024life}  recently adopted a Lagrangian potential vorticity (PV) perspective to understand the contribution of barotropic dynamics, nonlinear wave-eddy interactions and  moist processes to GB onset and persistence. More specifically, moisture was shown to influence both the initiation and maintenance of GB through radiative effects and latent heat release, possibly explaining its longevity relative to blocking in other regions of the North Atlantic \citep{Wallace1981teleconnections,pfahl2015importance,hauser2024life}. Beside investigating onset and maintenance mechanisms, \citet{hauser2024life} also classified GB into upstream and retrograding blocking regimes, two classes of blocks that differ in terms of onset pathway. The former originates upstream of Greenland while the latter follows a retrograding trajectory with respect to the flow. Given that moist diabatic processes act throughout the whole GB trajectory, they are often underestimated using Eulerian frameworks focusing only on local, rather than remote, processes \citep[e.g.,][]{Teubler2023dry}, motivating a greater reliance on Lagrangian approaches when studying GB dynamics.

Given the limitations associated with conventional Eulerian blocking definitions \citep{hauser2024life} and the GB observed trends, in this study we analyze the recent summer GB increase through a Lagrangian tracking algorithm, trying to understand whether the backtracking of geopotential height anomalies---and the inclusion and consideration of remote influences on blocking onset---can provide additional insights on the extreme ice sheet melting that occurred in the past decades. Moreover, following the approach of \citet{hauser2024life}, we separately examine trends in upstream and retrograding blocks to assess whether either type is driving the recent increase in summer blocking frequency. We provide additional insights on the features of these two families of blocking by developing an original classification method based on the wave-breaking characteristics. 

The Lagrangian tracking framework used in this study is a Python-based algorithm called "blocktrack", already illustrated and used in \citet{filippucci2024impact}. Thanks to its design, blocktrack enables detailed analyses of blocking trajectories duration, spatial extent, intensity and wave-breaking characteristics. Here, we first apply this algorithm to the ERA5 reanalysis dataset (1940--2024) and subsequently to a set of CMIP6 simulations, both for the historical and future periods. By analyzing the historical CMIP6 simulations we build on the study by \citet{maddison2024missing}, who highlighted how CMIP6 and HIGHRESMIP models fail to capture the recent increase in summer Greenland blocking, as well as its variability, using a similar ensemble. In this study we introduce a novel separation into upstream and retrograding blocks and apply this framework to the analysis of blocking frequency time series. Moreover, we expand upon \citet{maddison2024missing} by analyzing projections and interpreting projected blocking trends within the developed framework.

The paper is structured as follows: in Section \ref{sec:gb_method} we present the data used, the methods adopted and a detailed description of the Lagrangian tracking algorithm, in Section \ref{sec:gb_reanalysis} we analyze ERA5 reanalysis data and in Section \ref{sec:gb_trend_CMIP6} we present the CMIP6 analysis, for both the historical and future runs. Lastly, in Section \ref{sec:gb_conclusions}, we discuss our main results.

\section{Methods}
\label{sec:gb_method}
\subsection{Data}

The analyses presented in this paper are based on the ERA5 reanalysis \citep{hersbach2020era5} and an ensemble of models from the Coupled Model Intercomparison Project Phase 6 (CMIP6) \citep{eyring2016overview}. 

ERA5 is used as an observationally constrained estimate of historical atmospheric blocking activity over Greenland from 1940 to 2023 during NH summer (June, July and August, hereafter JJA), using daily data at horizontal resolution of 2.5° $\times$ 2.5°, to ensure a homogeneous spatial resolution across the datasets used in this study. The spatial and temporal resolution were chosen to resolve daily atmospheric variability at synoptic scales and perform a computationally efficient analysis. The geopotential height at 500 hPa (Z500) and the 2-meter temperature are retrieved at a single level, while the meridional wind component and specific humidity are retrieved across nine pressure levels.

To evaluate how current climate models represent atmospheric blocking over Greenland, we use a large multi-model ensemble of historical coupled model simulations from CMIP6. The choice of the ensemble was dictated by the availability of daily Z500 data for both historical and SSP3-7.0 simulations. Specifically, we analyze the geopotential height at 500 hPa retrieved from 70 ensemble members derived from 10 different models, each run in multiple configurations. The spatial and temporal resolution is the same as for ERA5. All ensemble members are weighted equally, so models with more members--such as UKESM--have a proportionally greater influence on the results. The  consistency of our results with those reported by \citet{maddison2024missing} gives us confidence that the selected ensemble is representative of the larger CMIP6 ensemble for the purposes of the present analysis. We report here for completeness that while blocking representation has significantly improved across different generations of CMIP models, an under-estimation of blocking frequency across both the Atlantic and Pacific basins persists even in the latest generation of model (CMIP6) \citep{davini2020cmip3,schiemann2020northern,dolores2025role}. In this study we therefore take into account this limitation by removing the effect of climatological blocking frequency biases on our analysis, as detailed in the following sections.

For projections, we rely on a smaller ensemble from the same set of 10 CMIP6 models, totaling 29 ensemble members. While a large ensemble is essential for quantifying internal variability in the historical period, a smaller ensemble size is sufficient to estimate future forced trends. The future scenario considered here is the SSP3-7.0, a high emission Shared Socioeconomic Pathway leading to a radiative forcing of about 7 W/m$^2$ by 2100. A detailed list of the models used and the corresponding number of ensemble members is provided in Appendix A2.

\subsection{Blocktrack}
\label{blocktrack}

The analysis presented in this paper uses the novel Python package "blocktrack" first presented in \citet{filippucci2024impact} and released on Github as \citet{blocktrackrepo}. This Python package is a flexible framework that allows for the detection of blocks through multiple blocking indices, the tracking of atmospheric blocking events, the computation and characterisation of block trajectories and the application of non-destructive filters to subsample detected blocking events according to user-defined criteria. A detailed description of each aspect of the algorithm is provided below.

\subsubsection{Blocked grid points identification}
The first step is the detection of blocked grid points in a gridded dataset through the analysis of daily Z500. Two indices---hereafter referred to as instantaneous blocking indices (IBI)---are adopted in this study and implemented in the blocktrack package: the geopotential height gradient reversal index (DAV), based on \citet{davini2012bidimensional}, and the geopotential height anomaly index (GHA), adapted from \citet{woollings2018blocking}, in turn similar to the index introduced by \citet{dole1983persistent}. The use of multiple indices reflects the diverse manifestations of GB, which can appear as persistent omega-shaped ridges or as high-amplitude Rossby waves breaking in a clockwise (anticyclonic) or anticlockwise (cyclonic) direction. As different indices are sensitive to different blocking structures or stages, combining complementary approaches has become standard practice to ensure robust detection and interpretation \citep{woollings2018blocking}.

In contrast with the work by \citet{hauser2024life}, who adopted a hybrid index, integrating a weather regime approach and the detection of PV anomalies, here we use geopotential height-based indices. While this choice is primarily motivated by the variable availability in our datasets, it also provides an opportunity to assess whether different blocking indices yield a consistent separation between upstream and retrograding blocks.

The DAV index identifies grid points where a reversal of the meridional geopotential height gradient occurs, making it particularly effective for detecting Rossby wave-breaking events. More specifically, for each grid point the northward ($GHGN$) and southward ($GHGS$) Z500 gradient are computed as:
\begin{equation}
    GHGN(\phi_0,\lambda_0) = \frac{Z500(\phi_N,\lambda_0)-Z500(\phi_0,\lambda_0)}{\phi_N - \phi_0}
\end{equation}
\begin{equation}
        GHGS(\phi_0,\lambda_0) = \frac{Z500(\phi_0,\lambda_0)-Z500(\phi_S,\lambda_0)}{\phi_0 - \phi_S} 
\end{equation}
where $\lambda_0$ and $\phi_0$ represent the grid point longitude and latitude, respectively; $\phi_0$ and $\lambda_0$ range from 30° to 75°N and 0 to 360°E, respectively; $\phi_s = \phi_0 - 15°$ lat and $\phi_N = \phi_0 +15°$ lat. A grid point of coordinates $(\lambda_0,\phi_0)$ is flagged as blocked if:
\begin{equation}
    GHGS(\phi_0,\lambda_0) > 0 ; \, GHGN(\phi_0,\lambda_0) < -10 \;\mathrm{m} \,\mathrm{lat}^{-1}
\end{equation}
We adopt the correction of \citet{tyrlis2021reconciling} that relaxes the northward gradient threshold to:
\begin{equation}
    GHGN(\phi_0,\lambda_0) < 0 \;\mathrm{m} \,\mathrm{lat}^{-1}
\end{equation}
This modification is more appropriate for high-latitude blocking events, where a strong jet on the poleward side of the block is less common.

In contrast, the GHA index identifies blocked grid points where daily Z500 anomalies, computed relative to a 90-day pointwise running mean, are larger than $1.26 \sigma$ of the Z500 anomaly distribution. The threshold has been chosen as the 90th percentile of a normal distribution. This probability distribution is calculated aggregating all the anomalies over the latitudinal band from 45°N to 80°N. The threshold is applied uniformly to all grid points and evolves smoothly through the seasonal cycle thanks to the 90-day pointwise running average used for the anomalies definition, thereby excluding the long-term thermal expansion of the atmosphere. 

Regardless of the IBI used, the output is a 3D (time, latitude, longitude) diagnostic matrix where each grid point is flagged as blocked or not blocked. 

\subsubsection{Tracking and characterization}
The next step is identifying blocked grid points that correspond to the same blocking event. This procedure, which represents the core of the tracking, is performed in two steps: (i) blocked areas are identified independently for each day by selecting contiguous blocked grid points, and (ii) blocked areas on consecutive days are considered part of the same event if they share at least 50\% of grid points with the previous day's area.

During event tracking, several characteristics are computed for each blocking event: area (in km$^2$) of the block on each day, center-of-mass trajectory, center-of-mass daily displacement, intensity, wave-breaking index (hereafter WBI) and duration. The intensity is calculated as the Z500 anomaly relative to a 90-day mean centered on the block day and averaged over the blocked area. The WBI, following \citet{davini2012bidimensional}, is defined as:
\begin{equation}
    WBI(\lambda_0,\phi_0) = \frac{Z_{500}(\lambda_W ,\phi_S+7.5°) - Z_{500}(\lambda_E ,\phi_S+7.5°) }{\lambda_W-\lambda_E }
\end{equation}
where $\phi_S$ is defined as in the blocking index and $\lambda_W = \lambda_0 - 7.5°$ and $\lambda_E = \lambda_0 + 7.5°$. Negative (positive) WBI values identify anticyclonic (cyclonic) wave breaking. The WBI index is then averaged over the blocked area and a single value is assigned to each blocked day. Note that in the case of a ridge or an omega block, WBI values lie close to zero. As a result, a tracked diagnostic matrix is produced in which each blocked grid point is labeled according to a unique blocking event. Moreover, a Python dictionary stores all block characteristics, providing a flexible dataset for analysis.

\subsubsection{Filtering}
The final step is filtering events. Depending on the chosen IBI, several filters can be applied to retain only events on the synoptic scale that persist for several days. In this study we use the following filters: (i) for DAV, a minimum daily average area of $5 \times 10^5$ km$^2$ and a minimum persistence of 5 days; (ii) for GHA, a minimum area of $2 \times 10^6$ km$^2$ and a minimum persistence of 5 days. The thresholds were selected to ensure consistency with previous studies \citep{davini2012bidimensional,woollings2018blocking} and reflect the tendency of the GHA index to detect larger blocked areas than the DAV index. 
These filters are applied non-destructively: a new diagnostic matrix is created in which events failing the criteria are excluded, without recomputing the tracking. This approach allows multiple filtering options to be tested easily and facilitates sensitivity analyses. We also verified that threshold changes do not significantly affect our results. In fact, we repeated the analysis without applying any filters to identify blocking events \citep{maddison2024missing} and we found that the time series of the atmospheric blocking frequency anomaly remains largely unchanged. This suggests that the filtering effectively removes short-lived and spatially small atmospheric blocking-like circulation patterns, while preserving the large-scale variability of blocking.

The algorithm is computationally efficient, requiring approximately fifteen minutes to track all blocking events in the ERA5 dataset (84 years, daily, 2.5° $\times$ 2.5° horizontal spatial resolution) on a regular laptop (Apple M2 processor). Moreover, we specify that the algorithm is applied separately to each CMIP6 ensemble model member (both historical and SSP3-7.0), before performing additional ensemble analysis. Lastly, we mention that other IBIs are implemented in the blocktrack algorithm and are available in the Github repository even if not used in this study.

\subsection{Frequency and composite analysis}
\label{frequencyandcompERA5}

Once the blocking events are detected, we isolate those occurring over Greenland. To do so we define a box around Greenland, ranging from 70W--20W and 60N--80N (illustrated in Fig.~\ref{trajectories}), and select the events that have at least one grid point within this region for at least three consecutive days. We opt for this selection method as it represents a trade-off between obtaining a large enough set of blocking events and selecting blocking events sufficiently localized over Greenland. Moreover, the method is analogous to the one adopted by \citet{maddison2024missing}. In this way a list of GB events is generated, along with their center-of-mass trajectories, WBI, average spatial extent, and other characteristics. For each blocking event, a single representative value of WBI and area is obtained by averaging over the event’s full life cycle.

The blocking composites are computed in a block-centered reference frame, i.e. they are Lagrangian composites. More specifically, for each blocking event, we extract the 2D field of interest (e.g., 2-m temperature T2m, integrated vapor transport IVT and Z500) within a box centered 10 degrees south of the event's center of mass and spanning 50 degrees of latitude and 90 degrees of longitude. The southward offset of the composite has been introduced to better highlight the equatorward features of the blocking system, otherwise not included in the composite. The average of these event-centered fields defines the Lagrangian composite. This approach allows us to capture the characteristics and spatial patterns of blocks, even when events occur at different geographical locations. We also compute composites across the different stages of the blocking life cycle. For days preceding the identified blocking onset, the daily composite field is computed by centering the box on the initial center-of-mass position of the block trajectory.

In the following sections we present composites of Z500, T2m and IVT, defined as in \citet{barrett2020extreme}:

\begin{equation}
\text{IVT} = \frac{1}{g} \int_{1000\ \text{hPa}}^{200\ \text{hPa}} q v \, dp
\end{equation}
where $g$ is the gravitational acceleration on Earth, $q$ is the specific humidity, $v$  the meridional component of wind and $p$ is the vertical pressure coordinate. The meridional velocity and the specific humidity are available at daily temporal spacing and on nine pressure levels. When plotting the composites of IVT we show its standardized anomaly  [IVT$^*]_{\sigma}^b$ computed as
\begin{equation}
    [\mathrm{IVT}^*]_{\sigma}^b =  \frac{[\mathrm{IVT}^*]^b}{\sigma_{\mathrm{IVT}}}
\end{equation}
where $[\mathrm{IVT}^*]^b$ is the composite anomaly obtained by averaging the IVT anomaly over blocking days and $\sigma_{\mathrm{IVT}}$ is the standard deviation of the IVT field. The standardized anomaly is useful for highlighting regions where IVT anomalies exceed typical variability.

Lastly, the blocking frequency time series presented below are computed by inspecting the box centered over Greenland (70W--20W and 60N--75N). For each day, we define the blocking frequency as:

\begin{equation}
    F_d = N_{blocked} / N_{tot}
\end{equation}
where $F_d$ is the daily frequency, $N_{blocked}$ and $N_{tot}$ are, respectively, the number of blocked grid points belonging to the previously identified set of Greenland blocks and the total number of grid points contained in the inspected area. To obtain a time series we compute the monthly mean of the daily frequency and its standardized anomaly, defined as:

\begin{equation}
    A_\sigma = \frac{F_m - <F_m>}{\sigma_{F_m}}
\end{equation}
where $F_m$ is the monthly blocking frequency, and $<F_m>$ and  $\sigma_{F_m}$ are, respectively, the climatological mean and standard deviation of the monthly values for the period 1950--2000. When analyzing CMIP6 projections, the mean and standard deviation are computed over the entire available period (2015--2100). We therefore refer to this quantity as 'normalized blocking frequency' rather than 'standardized anomaly'. Both the standardized anomaly and the normalized blocking frequency are useful for comparing model variability while removing model biases in climatological blocking frequency, which can vary considerably among models \citep{davini2020cmip3}. Finally, we smooth the time series by applying a 10-year running average. 

\section{Characteristics of summer Greenland blocking in reanalysis}
\label{sec:gb_reanalysis}
\subsection{Blocking event trajectories}

\begin{figure}
    \centering
    \includegraphics[width=1\textwidth]{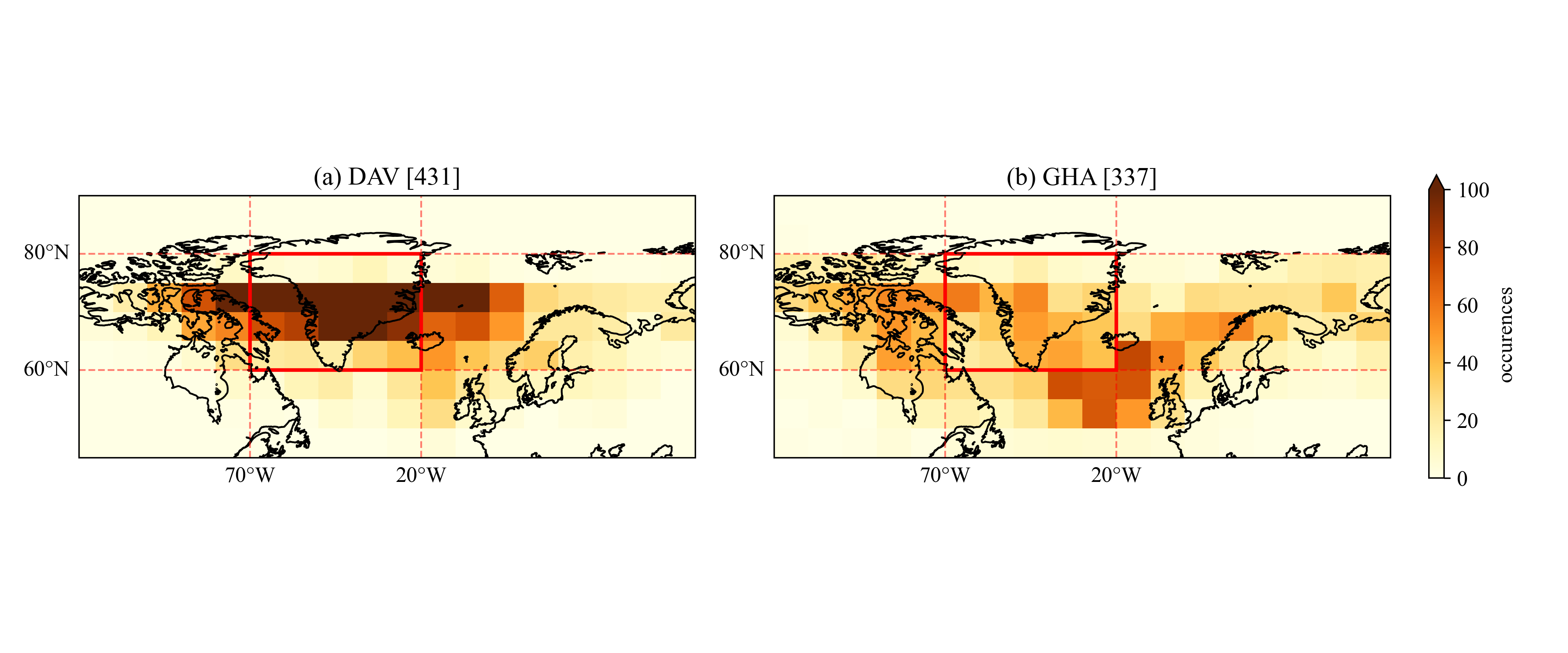} 
    \caption{GB events center-of-mass bidimensional distribution from ERA5 reanalysis (JJA). The two-dimensional histograms use 10°lon $\times$ 5°lat bins, and the value in each bin indicates how many times the center of mass of a GB event falls within that area. Panel (a) refers to the DAV index while panel (b) refers to the GHA index. The number of detected events crossing Greenland is shown in square brackets in each panel title. The red box identifies the region for GB event detection, as described in Section \ref{frequencyandcompERA5}}
    \label{trajectories}
\end{figure}

Fig.~\ref{trajectories} shows the center-of-mass positions of the blocking events crossing Greenland in ERA5 for JJA. The two panels present the center of mass derived using the DAV and GHA indices and tracked throughout the blocking life cycle, including days prior to the blocks' arrival over Greenland. 

The DAV index detects more blocking events compared to the GHA index, indicating that not all reversals of the meridional geopotential height gradient are associated with a large and persistent geopotential height anomaly. The events detected through the DAV index are more localized over Greenland throughout their life cycle (Fig.~\ref{trajectories}a), whereas those detected using the GHA index can travel longer distances (note that the map projection exaggerates distances at higher latitudes) and tend to form two clusters: one over northern Europe and one over northeastern Canada (Fig.~\ref{trajectories}b). This difference reflects the design of the two indices: GHA tends to detect earlier stages of blocking onset compared to DAV, since the ridge typically develops before Rossby wave breaking \citep{sousa2018european}. Moreover, the GHA index detects larger blocked areas as individual blocking systems, increasing the likelihood of event merging and splitting.

\begin{figure}
    \centering
    \includegraphics[width=1.0\linewidth]{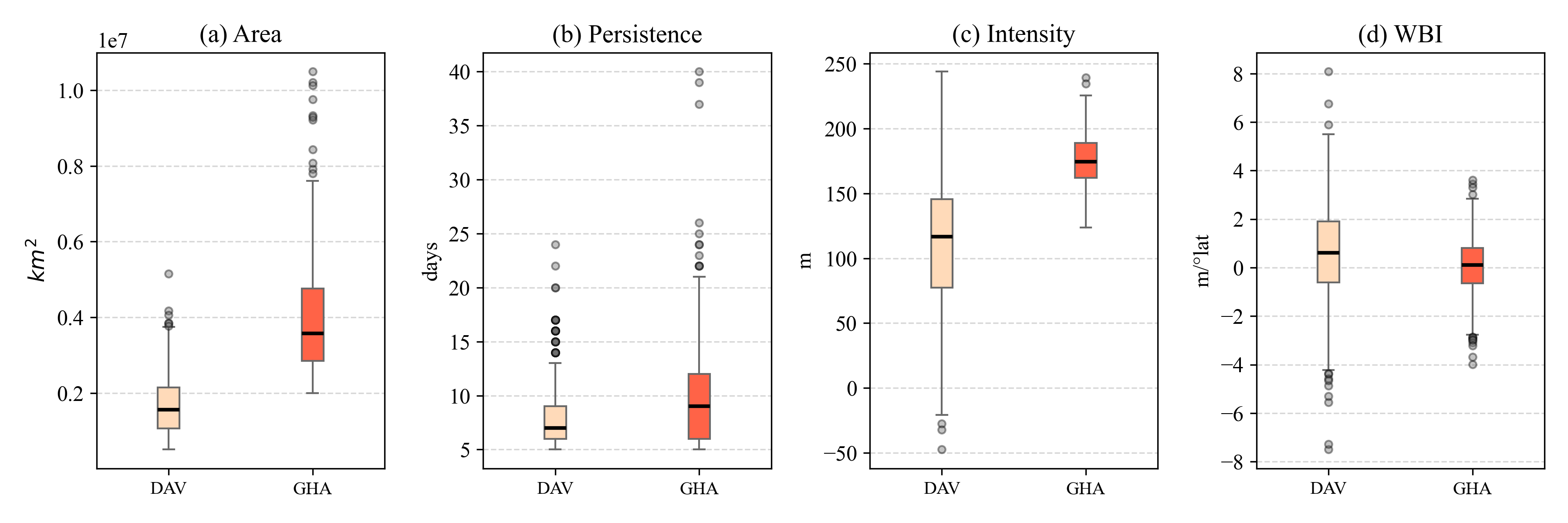}
    \caption{Box plots of Greenland blocking event characteristics in ERA5 reanalysis (JJA). Panel (a) shows event area [km$^2$], panel (b) shows event persistence [days], panel c) shows event intensity [m] and panel d) shows event WBI [m/$°$lat]. Orange and light orange indicate blocking events detected with the GHA and the DAV index, respectively. Black circles are outlier events.}
    \label{boxplot_all_events}
\end{figure}

As shown in Fig.~\ref{boxplot_all_events}, events detected using the GHA index are, on average,  about twice as large in area and tend to last longer than those detected using the DAV index. These characteristics are consistent with the broader center-of-mass distribution observed in Fig.~\ref{trajectories}. Greenland blocking events are associated with both positive and negative values of the WBI index (Fig~\ref{boxplot_all_events}d), indicating that wave breaking during GB events can be both cyclonic and anticyclonic. This holds true for both indices.

Moreover, the two hotspots of block trajectory frequency, one east and one west of Greenland, identified by the GHA index confirm the finding from \citet{hauser2024life} that GB events can originate from two preferred locations. Hereafter we adopt their same nomenclature, referring to blocking events preferentially residing to the east of Greenland as ''retrograding" blocks and those residing west of Greenland as ''upstream" blocks.


\subsection{Composite analysis}

\begin{figure}
    \centering
    \includegraphics[width=0.8\textwidth]{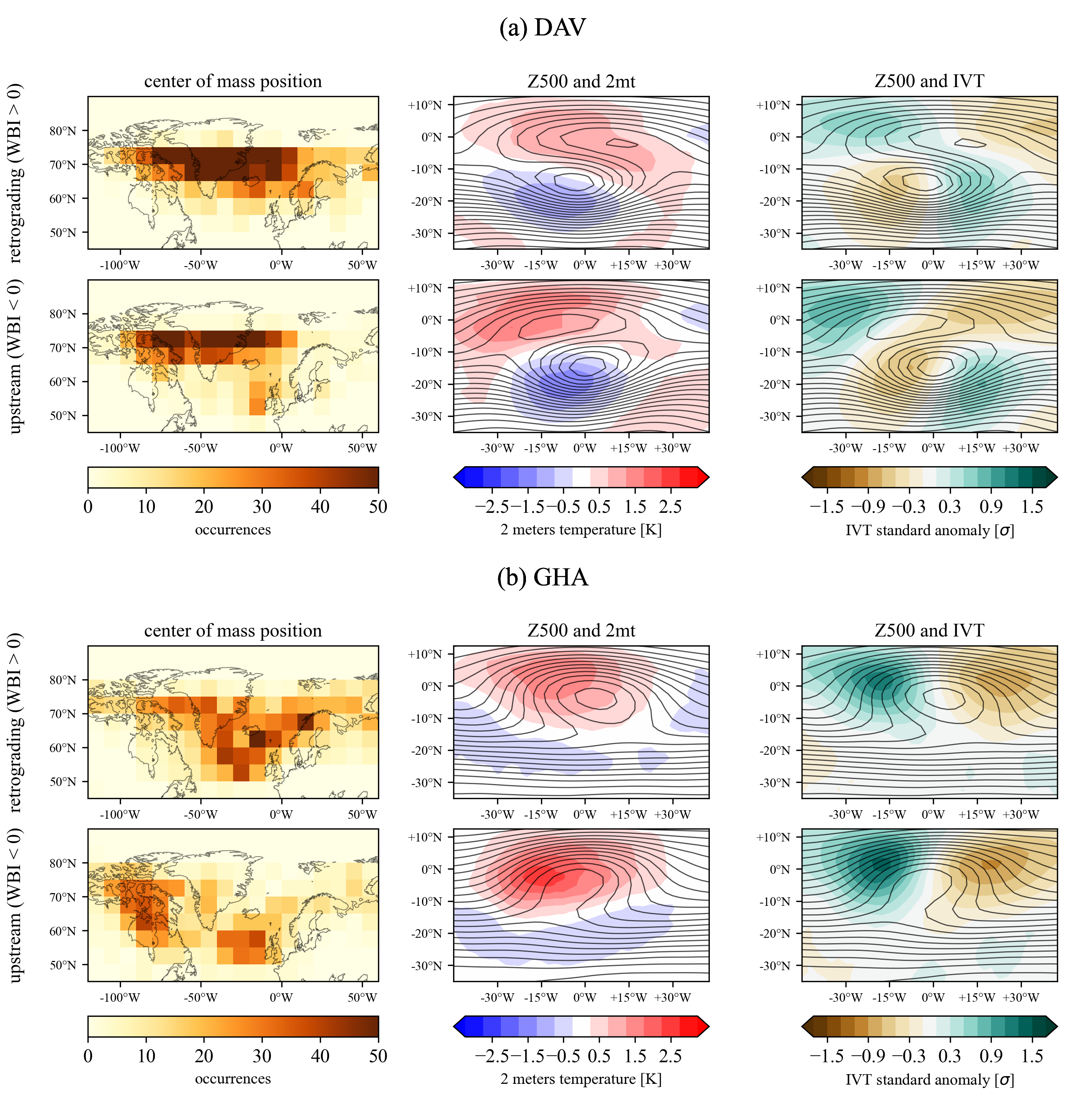} 
    \caption{Composites and center of mass positions of retrograding and upstream blocking events over Greenland from ERA5 reanalysis (JJA). The center-of-mass position plots are 2D histograms with 10°lon $\times$  5°lat bins, showing how often the center of mass of an upstream or retrograding block falls within each bin. Composite plots are plotted in the reference system of the block's center of mass (e.g. $x$-axis represents relative longitude). Black contours represent the 500-hPa geopotential height composite (Z500). Contours are plotted every 40 m. Shadings refer to the composite anomaly of temperature above the surface, T2m [K], and to the composite standard anomaly of IVT (see Methods for further details). Anomalies have been computed with respect to the seasonal mean. A Student’s t-test was performed to assess the significance of the plotted anomalies, with all shaded contours significant at the 95\% confidence level. Panel (a) shows blocking events detected through the DAV index and panel (b) blocking events detected through the GHA index.}
    \label{composites_ERA5}
\end{figure}    

We now separate the GB events into upstream and retrograding types and analyze their respective characteristics. To effectively separate blocking events in the two classes for both indices, we require an approach different from that of \citet{hauser2024life}, who based the classification on the geographical location of the blocking center of mass prior to its arrival over Greenland. In fact, for the DAV index, we do not find a clear separation into distinct geographical clusters. We therefore rely on the wave breaking characteristics of the blocks, which we find to be closely linked to the onset region and provide a consistent method for both the DAV and GHA indices.

The first column of Fig.~\ref{composites_ERA5} shows the center-of-mass density histograms of GB events for these two categories, with blocks labeled based on the sign of their WBI. In Fig.~\ref{composites_ERA5}b, based on the GHA index, it is evident that the WBI effectively distinguishes the two clusters of blocking events mentioned in the previous section: upstream blocks, originating west of Greenland, are associated with WBI$<0$ (anticyclonic breaking), whereas retrograding blocks, originating east of Greenland, are associated with WBI$>0$ (cyclonic breaking). We therefore adopt the same nomenclature for the blocking events detected through the DAV index, even though the spatial separation is less pronounced, because the WBI provides a consistent physical basis for classification. As stated in the previous sections, the GHA index captures earlier stages of the blocking life-cycle as expected, leading to a clearer separation into retrograding and upstream clusters. Lastly, a small secondary cluster of upstream GB centers of mass, identified by the GHA index in Fig.~\ref{composites_ERA5}b, is located in the eastern North Atlantic near the British Isles. These events likely correspond to anticyclonic Rossby wave breaking in northwestern Europe, which subsequently propagate over Greenland and are therefore dynamically distinct from the main upstream GB cluster. We excluded these events by reducing the size of the detection box (not shown), but this had little impact on the results presented here. Since the detection box reduction implies the exclusion of both this secondary cluster and a large fraction of the upstream block cluster, for the final analysis, we retained the larger box to enhance the statistical significance of our findings.

The second column of Fig.~\ref{composites_ERA5} (both panels a and b) shows the 2m-temperature and 500-hPa geopotential height composites of the so-identified blocking clusters. The Z500 contours  clearly depict the Rossby wave-breaking pattern associated with each block type for both indices. As expected, for $WBI<0$ ($WBI>0$ ) Rossby wave breaking occurs anticyclonically (cyclonically). For blocking events detected with the DAV index (Panel a) the 2m temperature anomalies exhibit a dipole pattern, with warm anomalies on the northern flank and cold anomalies on the southern flank, consistent with the ridge-trough circulation of the block. These anomalies arise from a combination of meridional advection of warm/cold air and radiative effects. In contrast, for blocks detected using the GHA index (Fig.~\ref{composites_ERA5}b) the dipole pattern is weaker, if not absent, and positive temperature anomalies are larger than in Fig.~\ref{composites_ERA5}a. These differences reflect the design of the indices: GHA captures positive geopotential height anomalies and earlier stages of block development, while DAV  does not require strong positive geopotential height anomalies and is often linked to weaker anticyclonic circulation. This distinction is also consistent with the differences in mean intensity shown in Fig.~\ref{boxplot_all_events}.

Moisture transport plays a key role in the radiative and advective processes associated with blocking onset and impacts  \citep{pfahl2015importance,liu2015extreme,hauser2024life,dolores2025role}. To highlight this effect, we plot composites of the IVT standardized anomaly (see the Methods section for further details). In both DAV (panel a) and GHA (panel b) composites, the IVT standard anomaly closely follows the meridional wind pattern, with negative anomalies upstream and positive anomalies downstream of the block. This pattern is consistent with the geostrophic approximation, where Z500 contours represent the streamfunction of the horizontal flow. Comparing the two families of blocking we find that the IVT standard anomaly is generally stronger for upstream blocks than for retrograding blocks for both indices. Moreover, anomalies are particularly large upstream and northwest of the block, an area corresponding to the Labrador sea and the Baffin Bay, identified by previous work as crucial for blocking intensification due to moisture advection \citep{liu2015extreme}. However, moisture advection alone is not sufficient to drive blocking amplification---latent heat release by ascending air streams must also occur. As a proxy for this process, we examine precipitation composites during blocked days, presented in the Supplementary material (Fig.~S1), which reveals enhanced precipitation to the northwest of upstream blocks, consistent with the increased meridional water vapour transport identified above. Notably, the synoptic circulation associated with upstream blocks intensifies the advection of moisture toward the northeast, a configuration compatible with the development of foehn winds northeast of the Greenland ice sheet, a phenomenon that \citet{mattingly2023increasing} associated to extreme ice melt. We will return to this point in the Conclusions.

We further evaluate if upstream and retrograding blocks differ systematically in area, persistence, or intensity, for both indices (not shown). No statistically significant differences are found, likely because the variability within each block type is large relative to the differences between types (see the boxplot in Fig.~\ref{boxplot_all_events}).

\subsection{Time evolution of Greenland blocking}

\begin{figure}
    \centering
    \includegraphics[width=0.8\linewidth]{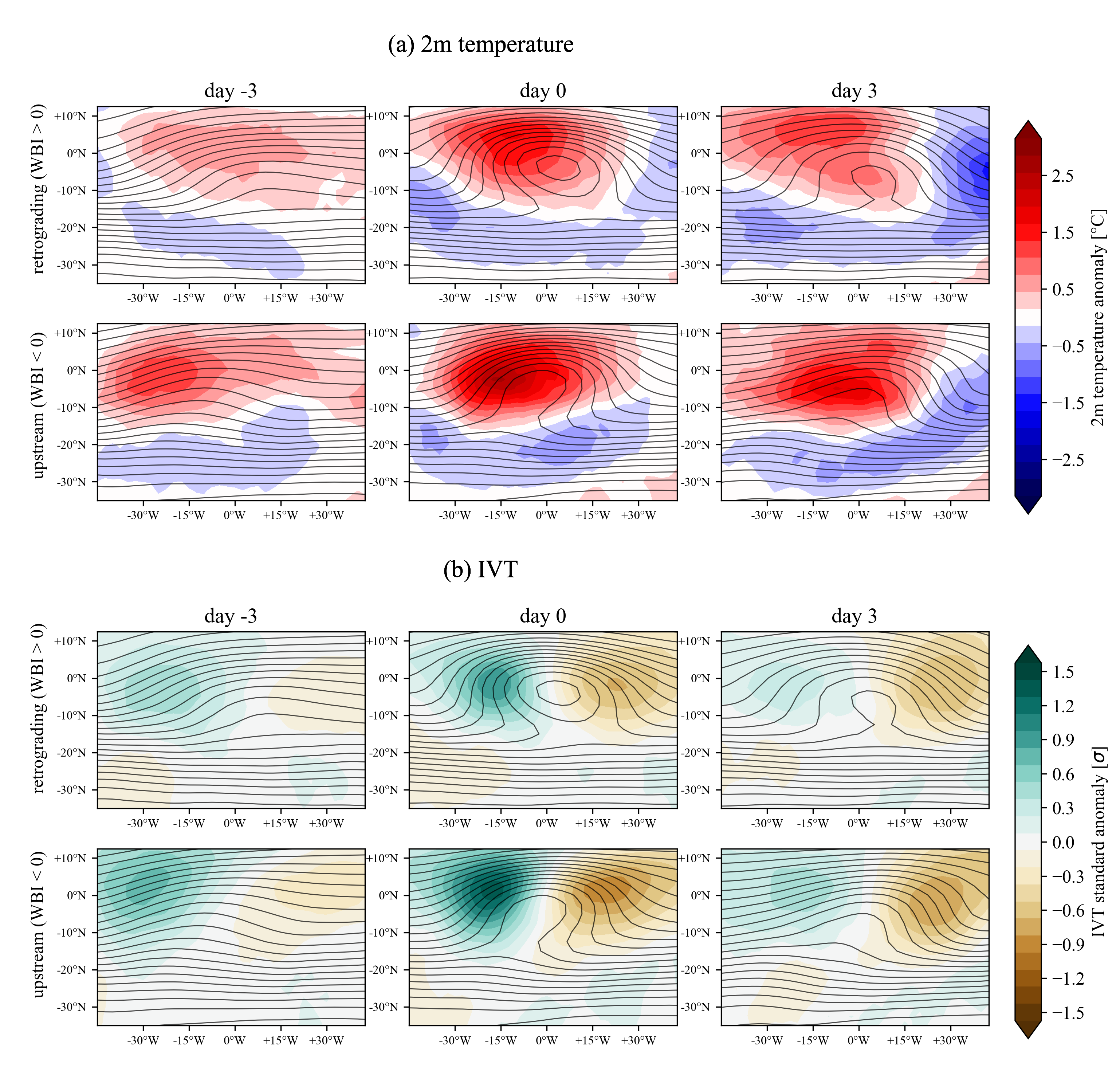}
    \caption{Composites of the time evolution of retrograding and upstream blocking events over Greenland from ERA5 reanalysis (JJA) detected through the GHA index. Composite plots are plotted in the reference system of the block's center of mass. Black contours represent the 500-hPa geopotential height composite (Z500). Contours are plotted every 40 m. Shadings refer to the composite anomaly of temperature above the surface (2 meters temperature T2m [K]) (panel a) and to the composite standard anomaly of IVT (see methods for further details) (panel b). Anomalies have been computed with respect to the seasonal mean. A Student’s t-test was performed to assess the significance of the plotted anomalies, with all shaded contours significant at the 95\% confidence level. }
    \label{evolution_composites}
\end{figure}
In Fig.~\ref{evolution_composites} we show composites of upstream and retrograding blocks over Greenland during different stages of their life cycles, as detected by the GHA index. See Supplementary Material for a similar analysis based on the DAV index showing comparable features (Fig.~S2). Again, we see that temperature anomalies are larger for upstream blocks. In both cases, the rise in temperature starts before the blocking when the ridge structure starts developing. This feature is particularly evident for the upstream blocks. IVT standard anomalies are again larger for upstream blocks than retrograding blocks, especially before and during the blocking onset. This fact suggests a link between the fast temperature rise and the moisture advection, which likely contributes to the development of the block.

Seasonal analyses reveal pronounced variations in the relative frequency of upstream and retrograding blocks. Upstream blocks are less frequent in winter, reflected in a shift of the WBI distribution toward positive values (see Appendix and Fig.~\ref{WBI_annual_cycle} for a more detailed analysis). This suggests a strong seasonal dependence in the underlying blocking dynamics, likely linked to the poleward migration of the polar jet stream in summer and its equatorward shift in winter \citep{woollings2010variability}, as well as to differences in insolation and the considerably different ocean surface energy balance around Greenland during summer. It has been shown that the position and strength of upper-level winds and their interaction with orography is crucial for setting the frequency and location of blocking onset \citep{ji1983numerical,berckmans2013atmospheric,davini2021orographic}. The seasonal migration of the polar jet stream toward Greenland's latitudes together with the fact that the Greenland ice dome can reach altitudes of 3200 m \citep{hawley2020greenland} is therefore likely to be at least partly responsible for the seasonal variations of Greenland blocking. Moreover, the separation into upstream and retrograding blocks might be facilitated by Greenland orography, which may interact with the low-level flow by altering the evolution of the synoptic flow (for example, via latent heat release by ascending air motion or lee cyclogenesis). Regarding this latter mechanism, the presence (or lack thereof) of sea ice over the waters around Greenland may play a crucial role, influencing the moisture exchange between atmosphere and ocean. We will return to this point in the Conclusions section.

To summarize, we find that Greenland blocking tends to occur via two distinct pathways, termed upstream and retrograding following \citet{hauser2024life}, which feature distinct characteristics. Upstream blocks, originating in Northern Canada and associated with anticyclonic Rossby wave breaking, are linked to stronger temperature anomalies compared to retrograding blocks, originating in Northern Europe and associated with cyclonic Rossby wave breaking. Importantly, upstream blocks are accompanied by stronger meridional moisture fluxes, which have been argued by previous work to play an important contribution to both blocking onset and amplification.

\section{Trends in reanalysis and CMIP6 models }
\label{sec:gb_trend_CMIP6}
\subsection{ERA5 and CMIP6 historical runs}

\begin{figure}
    \centering
    \includegraphics[width=1.0\linewidth]{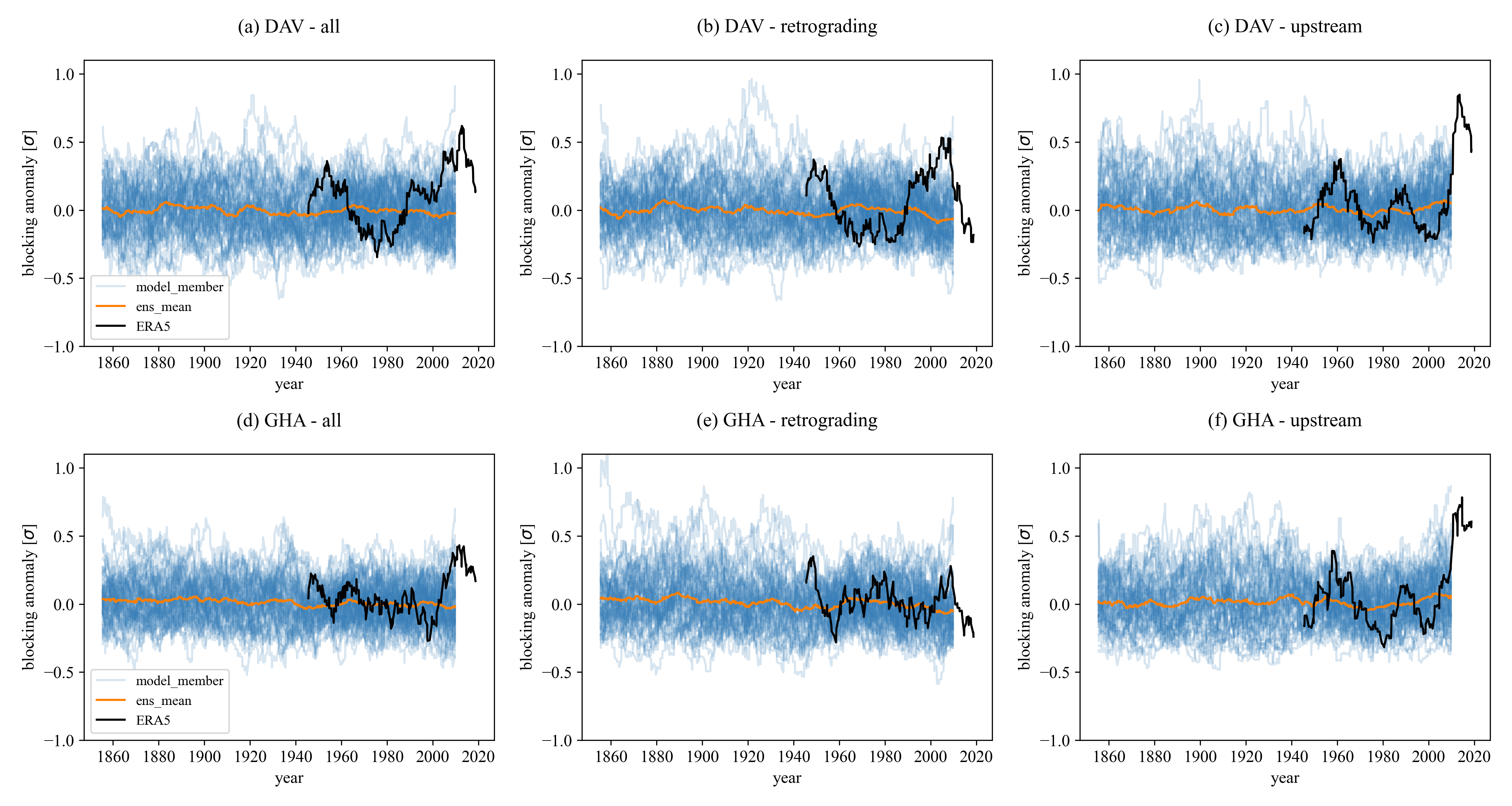}
    \caption{Summer (JJA) blocking standard anomaly time series (see  Methods  for further details on its definition). In all panels, black lines refer to ERA5 reanalysis dataset, blue lines refer to CMIP6 ensemble members and orange lines refer to the ensemble average. The top row (a--c) shows blocking anomalies detected through the DAV index, while the bottom row (d--f) shows anomalies detected through the GHA index. The left column (a,d) refers to all blocking events crossing Greenland, the center column (b,e) refers to retrograding blocks and the right column (c,f) refers to upstream blocks.}
    \label{frequency_trend_historical}
\end{figure}

The timeseries of summer blocking frequency anomaly from ERA5 and from an ensemble of CMIP6 models is shown in Fig.~\ref{frequency_trend_historical}. The anomaly is computed by taking the monthly blocking frequency for each model/reanalysis product, subtracting the mean, dividing by its standard deviation and ultimately performing a 10-year running mean (for further details see the Methods section).

Both DAV and GHA identify an increase in GB frequency in the first two decades of the 21st century in ERA5 (Fig.~\ref{frequency_trend_historical}a--d), with a peak in August 2012 for DAV and June 2013 for GHA (not shown). Even though some simulations reach a similar or even higher blocking anomaly, it is evident how the ERA5 maximum sits at the upper margin of the anomaly distribution of the model ensemble for both indices, as previously identified by \citet{hanna2016greenland} and \citet{maddison2024missing}. We now construct similar time series for upstream and retrograding blocking separately (Fig.~\ref{frequency_trend_historical}b--f). This separation highlights an important and so far unreported result: the largest contribution to the recent summer blocking frequency increase comes from upstream blocks, with the largest increase occurring in the second decade of the 21st century (Fig.~\ref{frequency_trend_historical}c, f). While true for both indices, this trend is more evident in the DAV index-based time series.

In Fig.~\ref{frequency_trend_historical} we analyze standardized anomalies to highlight trends and variability. However, because the mean is subtracted in this process, this figure does not reveal whether the climatological partitioning of upstream versus retrograding blocks is well represented by the CMIP6 models. A separate analysis performed through the absolute frequency rather than blocking anomaly reveals that CMIP6 models do capture well the separation into upstream and retrograding summer blocks in the 1950--2000 reference period (see Supplementary Material, Fig.~S3). Approximately one third of the blocks are classified as upstream and two thirds as retrograding for both CMIP6 models and reanalysis, even though the absolute frequency is slightly underestimated by the CMIP6 models. 

To quantify the magnitude of the recent observed peak, we compare the maximum blocking frequency anomaly in ERA5 against the distribution of maxima from the CMIP6 ensemble. By frequency maximum we mean the maximum anomaly of blocking frequency in a month, relative to the climatology of monthly blocking frequency for all JJA months in the historical record.

\begin{figure}
    \centering
    \includegraphics[width=1.0\linewidth]{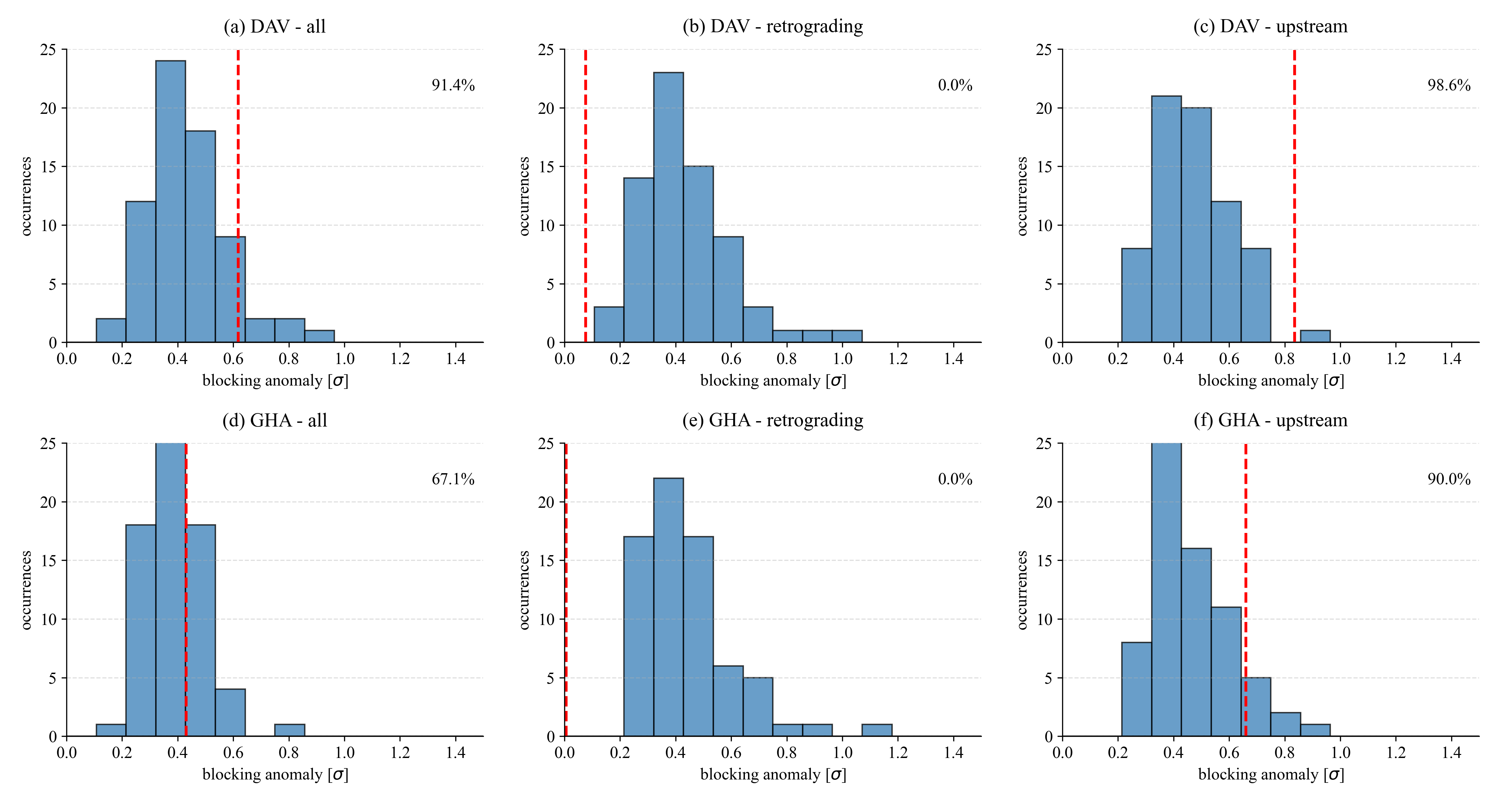}
    \caption{Distributions of frequency maxima in the CMIP6 models ensemble and comparison with the recent Greenland blocking increase in ERA5. The blue histograms show the distribution of the all-time maximum anomaly found in each model's 84-year window. The vertical red line represents the recent ERA5 summer Greenland blocking maximum. For the upstream and retrograding panels, the ERA5 line represents the anomaly value observed during the month of the total blocking maximum. In the bottom right of each plot we show the percentile corresponding to the ERA5 frequency value. Note that the distribution maxima and the ERA5 maximum have been computed taking into account the frequency timeseries smoothed through a 10 years rolling mean. Top row (a--c) shows the analysis conducted with the DAV index, while the bottom row (d--f) shows the analysis based on the GHA index. The left column (a,d) refers to all  blocking events crossing Greenland, the center column (b,e) refers to retrograding blocks and the right column (c,f) refers to upstream blocks}
    \label{frequency_distribution_historical}
\end{figure}

To obtain the CMIP6 distribution of maxima, for each of the 70 simulations we extract the maximum anomaly occurring within an 84-year time window (1930--2014), a time span chosen to match the ERA5 record length (1940--2024). This is done independently for the total, retrograding, and upstream blocking time series, with the resulting distributions of model maxima based on both DAV and GHA indices plotted in Fig.~\ref{frequency_distribution_historical}. 

For the total blocking frequency (Fig.~\ref{frequency_distribution_historical}a,d), the recent ERA5 maximum lies above the median (50th percentile) of the CMIP6 distribution, indicating that the reanalysis exhibits larger blocking anomalies than most models. This is especially evident for the DAV index (91.4th percentile), and also holds true for the GHA index (67.1th percentile), confirming results in \citet{maddison2024missing}. 

We then analyze the composition of  the recent ERA5 blocking anomaly maximum, separating the contributions from upstream and retrograding blocks (Fig.~\ref{frequency_distribution_historical}b,e,c,f). This is done by inspecting the upstream and retrograding blocking frequency time series in the same month of the observed blocking frequency maximum (August 2012 for the DAV index and June 2013 for the GHA index). Note that the monthly time series is smoothed by applying a 10-year running mean (for further details see the Methods section). Crucially, while retrograding blocks do not exhibit notable anomalies during the observed ERA5 peak, which falls well below the typical model maxima, a very high value of upstream block frequency anomaly is observed for both the DAV and GHA indices. For the DAV index, the ERA5 upstream blocking anomaly is overcome by just 1 of the 70 simulations composing the CMIP6 ensemble (98.6th percentile), implying that CMIP6 models likely underestimate the variability of upstream blocking. Results are consistent for the GHA index, where the percentile value remains high (90th percentile), confirming that the recent increase in GB is indeed driven by a period of extreme upstream blocking frequency.

Finally, we assess whether the physical characteristics of upstream and retrograding blocks are correctly captured by CMIP6 models. By comparing the distributions of center-of-mass position, area, persistence, WBI, and intensity between the CMIP6 ensemble and ERA5, we find that the models reproduce the observed blocking features with high fidelity. In fact, the distributions of the GB event characteristics are almost identical in ERA5 and the historical CMIP6 ensemble. This is especially true for the DAV index, for which no appreciable differences are found. For the GHA index, however, the models slightly underestimate the persistence and intensity of both block types. A complete set of characteristics histograms is provided in the Supplementary Material (Figs.~S4 and S5).

We do not present composite plots of upstream and retrograding blocks for the CMIP6 ensemble, as a detailed model-by-model composite analysis is impractical for the large ensemble size considered here and falls outside the scope of this paper. However, the fact that blocking characteristics are well captured gives us confidence that the fundamental processes driving the onset and maintenance of both upstream and retrograding blocks are well represented. This conclusion is further supported by results in \citet{luu2024greenland}, who analyzed a smaller subset of CMIP6 and HIGHRESMIP models. They found that while models fail to capture the recent increase in frequency, they reproduce the spatial structure and features of blocking reasonably well.

\subsection{CMIP6 projections}

\begin{figure}
    \centering
    \includegraphics[width=1.0\linewidth]{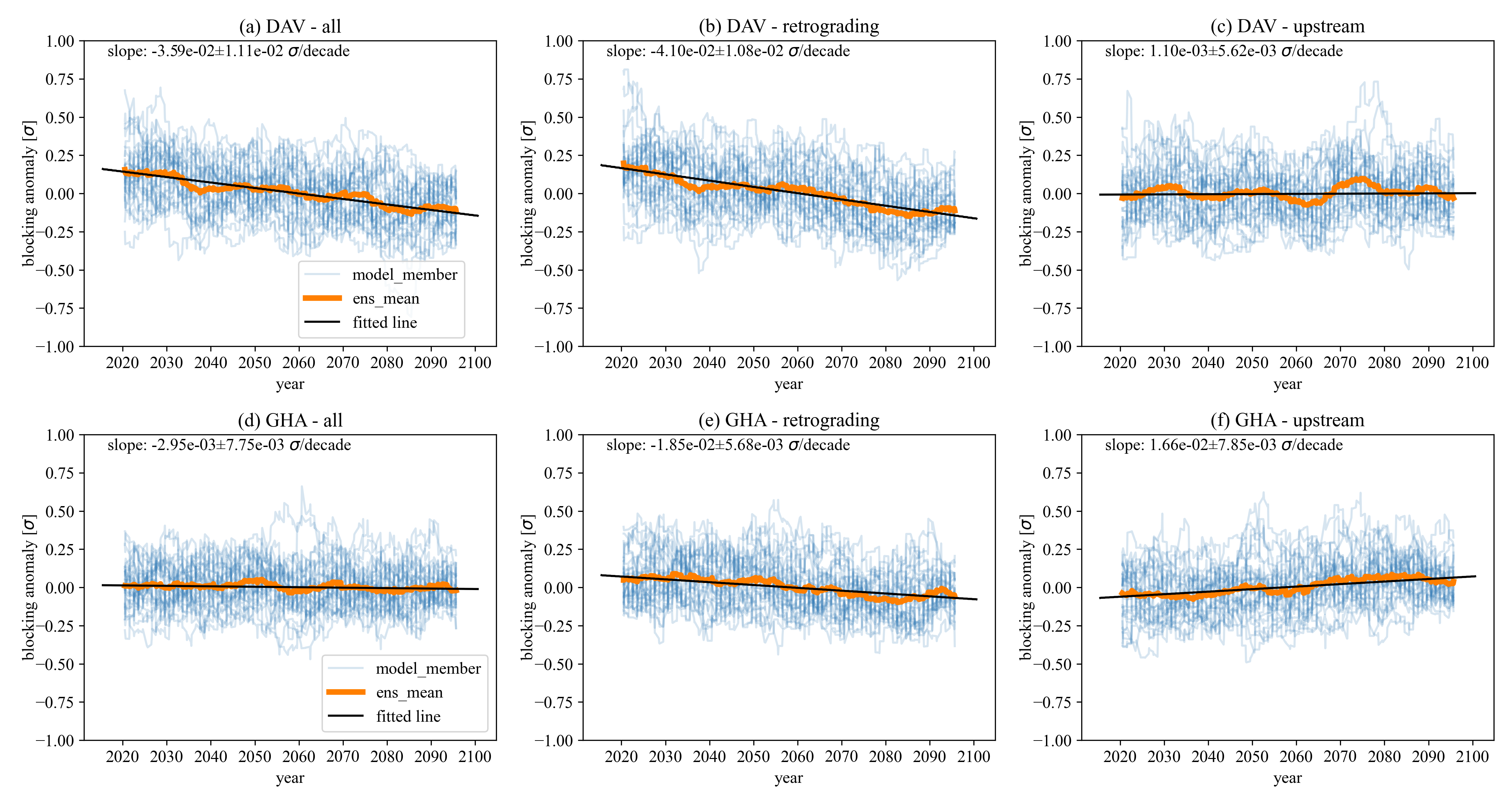}
    \caption{Future blocking standard anomaly timeseries (see Methods  for further details on its definition). In all panels, blue lines refer to CMIP6 ensemble members, orange lines refer to the ensemble average and the black lines refer to the linear regression of the ensemble average. On top of each plot we report the linear regression parameters. Top row (a--c) shows blocking anomalies detected through the DAV index, while bottom row (d--f) shows anomalies detected through the GHA index. Column (a,d) refers to all blocking events crossing Greenland, column (b,e) refers to retrograding blocks and column (c,f) refers to upstream blocks.}
    \label{frequency_trend_future}
\end{figure}

In Fig.~\ref{frequency_trend_future}, we present projected future trends in normalized summer GB frequency. The normalized blocking frequency is calculated similarly to the blocking anomaly, except that the standard deviation and mean are derived over the entire simulation period (2015--2100) (see Methods for details).

Fig.~\ref{frequency_trend_future}(a,d) show the projected evolution of total normalized blocking frequency over Greenland in summer, as computed using the DAV and GHA indices. A clear decreasing trend emerges when using the DAV index (Fig.~\ref{frequency_trend_future}a); the GHA index (Fig.~\ref{frequency_trend_future}d) instead does not exhibit any significant change over time. The trend significance is evaluated by comparing the slope of the least squares linear regression with its associated error; we consider it significant when the slope’s magnitude exceeds twice the standard error (i.e., lies outside the $2\sigma$ range). These projections provide a useful comparison to \citet{delhasse2021brief} (hereafter D21), who analyzed projections of summer Greenland blocking in a CMIP6 model ensemble. Our analysis diverges from D21 in two key aspects: first, we employ the SSP3-7.0 scenario (versus SSP5-8.5 in D21); second, while D21 used a GBI based on average Z500, we analyze trends using both the DAV and the GHA indices. 

Discriminating the total blocking signal into upstream and retrograding components reveals distinct and so-far unreported dynamical trends. The projected decline in total GB frequency is driven primarily by a significant reduction in retrograding blocks (Fig.~\ref{frequency_trend_future}b,e). While this reduction is captured by both indices, it is most pronounced in the DAV index (Fig.~\ref{frequency_trend_future}b). In contrast, projections for upstream blocking differ between the two definitions (Fig.~\ref{frequency_trend_future}c,f). The GHA index indicates a marked increase in upstream blocking---consistent with recent anomalies in reanalysis---whereas the DAV index shows no significant trend. This explains why the total GHA frequency remains steady: the projected rise in upstream blocks effectively compensates for the reduction in retrograding blocks. Such compensation is not seen in the DAV index. 

In summary, we find another important and so-far unreported result: the projected decrease of atmospheric blocking frequency (already identified by D21) can be attributed to retrograding blocks, while upstream blocks remain constant or increase, depending on the index used. 
 
Lastly, we compare the features of GB events (persistence, area, WBI and intensity) between historical simulations and projections (not shown). For both indices we find a small shift of the WBI distribution toward negative values in the projections, coherent with a decrease of retrograding blocks and an increase of upstream blocks. Other characteristics remain relatively unchanged. Please see the associated histograms in the Supplementary Material (Figs.~S6 and S7).

\section{Conclusions}  
\label{sec:gb_conclusions}

In this study, we have analyzed the characteristics, historical trends, and projections of summer GB using a novel Lagrangian framework applied to the ERA5 reanalysis and an ensemble of CMIP6 models.

First, we examined the geographical distribution, trajectories and characteristics of GB events. We found that GB can originate through both anticyclonic and cyclonic Rossby wave breaking. This distinction is found to be associated with two main categories of GB: retrograding blocks, typically linked to cyclonic wave breaking and propagating from northern Europe, and upstream blocks, associated with anticyclonic wave breaking and originating in northeastern Canada. The geographical separation between these two categories is especially clear when using an anomaly-based blocking index, such as GHA, which better captures the earlier stages of the blocking life cycle and diagnoses events with larger spatial extent. A similar separation has been previously adopted by \citet{hauser2024life}, who first introduced the term 'upstream' and 'retrograding' blocks. However, while \citet{hauser2024life} separate events based on the geometric center of mass position prior to onset, here this classification naturally emerges from the mean WBI value during the blocking life cycle.

To further investigate the nature of these blocks, we constructed Lagrangian composites of 2-meter temperature, Z500, and IVT for both retrograding and upstream blocks. Regardless of the blocking index used, we find that upstream blocks are typically associated with stronger temperature anomalies and enhanced meridional moisture transport compared to retrograding blocks. Notably, these thermodynamic anomalies begin developing up to three days before the blocking onset, suggesting a critical preconditioning role of moisture advection. Moreover, composite plots of the total precipitation during and before blocking onset show similar anomalies (see Supplementary Material, Fig.~S1), implying latent heat release by the advected moist air masses. These results agree with previous studies by \citet{pfahl2015importance} and \citet{hauser2024life}, which highlighted the influence of latent heat and moist processes on blocking formation and persistence. Specifically, \citet{pfahl2015importance} argue that latent heat release associated with the upward transport of air from low levels contributes significantly to the upper-troposphere blocking anticyclone. Similarly, \citet{hauser2024life} found that the amplification of Greenland blocking through moist processes occurs earlier in upstream blocks than in retrograding blocks, suggesting a more dominant role of moisture in upstream block development. Our composite analysis confirms these findings, revealing that thermodynamic anomalies and moisture advection into the high latitudes develop several days prior to the onset of upstream blocking events.

Importantly, our analysis shows that the observed increase in Greenland blocking frequency during the early 21st century is primarily driven by an increase in upstream blocks. This is a novel result, although recent literature has increasingly emphasized the importance of processes occurring west of Greenland. For instance, \citet{preece2023summer} identified North American snow cover spring anomalies and Arctic Amplification as key drivers of early summer Greenland blocking anomalies, while \citet{beckmann2025summer} noted that the Atlantic Multidecadal Variability (AMV) and Arctic Amplification can alter the Labrador sea surface temperature, thereby affecting local atmospheric circulation. Moreover, \citet{beckmann2025summer} confirmed the findings of \citet{preece2023summer} regarding the crucial role of spring snow cover in Northern Canada. These phenomena can induce warm temperature anomalies west of Greenland, triggering a stationary Rossby wave that favors the anchoring of eastward propagating ridges over Greenland, consistent with our findings. Ultimately, the link between moisture and blocking connects our findings to broader research on extremes. For example, \citet{barrett2020extreme} demonstrated the role of moisture transport over the Labrador Sea in amplifying extreme blocking events, such as the ones that occurred in the years 2012 and 2019 \citep{tedesco2020unprecedented}. Similarly, \citet{scholz2024atmospheric} found a strong correlation between major atmospheric river events and high-latitude heatwaves, particularly over northeast Canada. Our finding that upstream blocks are intrinsically linked to high-latitude moisture advection supports the hypothesis that the impact of such atmospheric rivers may increase in a warmer, moister atmosphere.

Lastly, we evaluated future trends in Greenland blocking using CMIP6 projections under the SSP3-7.0 warming scenario. While retrograding blocks are projected to decrease significantly---explaining the declining total blocking frequency reported by \citet{delhasse2021brief} and \citet{hanna2016greenland}---upstream blocks show an increasing trend, particularly when assessed through the GHA index. This index detects anticyclonic anomalies rather than atmospheric flow patterns, making it more sensitive to changes in blocking intensity and persistence: more intense Z500 anomalies tend to produce longer-lived and spatially larger blocking events. Under enhanced warming, moisture advection into the blocking centre is expected to intensify these anomalies, which would manifest more clearly in the GHA index than in the DAV index. Given that CMIP6 models already underestimate both the historical magnitude and trend of GB frequency \citep{maddison2024missing}, and that the GHA index is precisely the one capturing this intensity-driven component, our findings suggest that models point to---or at least do not rule out---a future increase of upstream GB.

The pronounced seasonality of GB may offer a useful framework for interpreting both the observed increase in GB frequency and its projected future evolution. As anticipated in the Results section and illustrated in the Appendix (Section~\ref{appendixA}), upstream blocks occur predominantly in summer, whereas retrograding blocks are more common in winter. During winter, the eddy-driven jet stream sits at relatively low latitudes and solar radiation reaches Greenland for only a few hours per day; in summer, insolation is stronger and the eddy-driven jet stream is weaker and displaced further north, extending to Greenland's latitude. These two fundamental seasonal differences may influence blocking development over Greenland through at least two distinct mechanisms. (i) The Greenland ice dome, reaching elevations of up to 3200 meters \citep{hawley2020greenland}, can interact with upper-level winds to generate a stationary Rossby wave response, thereby shifting the preferred regions of blocking onset \citep{held1990orographic,held2002northern,woollings2018blocking,nakamura2018atmospheric}; (ii) stronger solar radiation, higher atmospheric temperatures, and reduced sea ice cover over the Labrador Sea and the Baffin Bay may collectively increase water vapor content to the west of Greenland, intensifying upstream blocks. More precisely, as shown in Fig.~\ref{composites_ERA5}, the synoptic circulation associated with upstream blocks acts to tilt the eddy-driven jet stream northeastward, advecting moist air from the Labrador Sea and the Baffin Bay further north and up Greenland's orography. This configuration may trigger a stau-foehn effect, enhancing precipitation on the northwestern flank of the block (Fig.~S1 in the Supplementary Material) and further amplifying low-level warming within the blocking center. The contribution of foehn winds to heat extremes over northeast Greenland has been established by \citet{mattingly2023increasing}, who demonstrated that a high percentage of exceptional ice melt events over the last few decades were linked to anomalously high atmospheric river activity over western Greenland and associated foehn conditions in the northeast. 

Greenland blocking projections can be interpreted in light of the seasonal differences discussed above, with summer-like conditions becoming more probable as global temperatures rise. For instance, if global warming were to alter the position of upper-level winds, increasing the likelihood of summer-like jet stream positioning with respect to Greenland orography, we might expect a corresponding change in the relative frequency of upstream versus retrograding blocks, especially in early summer. While the future response of the North Atlantic jet stream remains a subject of debate, many studies suggest a poleward shift in summer \citep{simpson2014diagnosis,shaw2016storm,chen2020sensitivity,zhou2022seasonally}, a mechanism consistent with our hypothesis. Moreover, warmer sea surface temperatures and sea-ice free conditions in the Labrador sea and the Baffin Bay may enhance the advection of moist air over Greenland, further amplifying the blocks. These conditions may arise as a consequence of global warming, Arctic amplification or AMV. Ultimately, confirming this causal chain would require targeted numerical experiments designed to separate thermodynamic and dynamical effects and a more granular analysis of local circulations and individual blocking events, tasks that could motivate future work.

In conclusion, the observed and projected rise in blocking frequency may be linked to jet stream position shifts, sea-ice reduction over the Labrador Sea and Baffin Bay, reduced spring snow cover in North America, and intensified moisture transport and atmospheric rivers over the North Atlantic. Our Lagrangian analysis offers a new perspective on these phenomena, suggesting that future trends in Greenland blocking may hinge on the behavior of its upstream component.

\section{Appendix}

\let\oldthefigure\thefigure

\renewcommand{\thefigure}{A3.\arabic{figure}}
\setcounter{figure}{0} 

\subsection{Seasonal cycle of Greenland blocking frequency and WBI}    
\label{appendixA}

\begin{figure}
    \centering
    \includegraphics[width=1.0\linewidth]{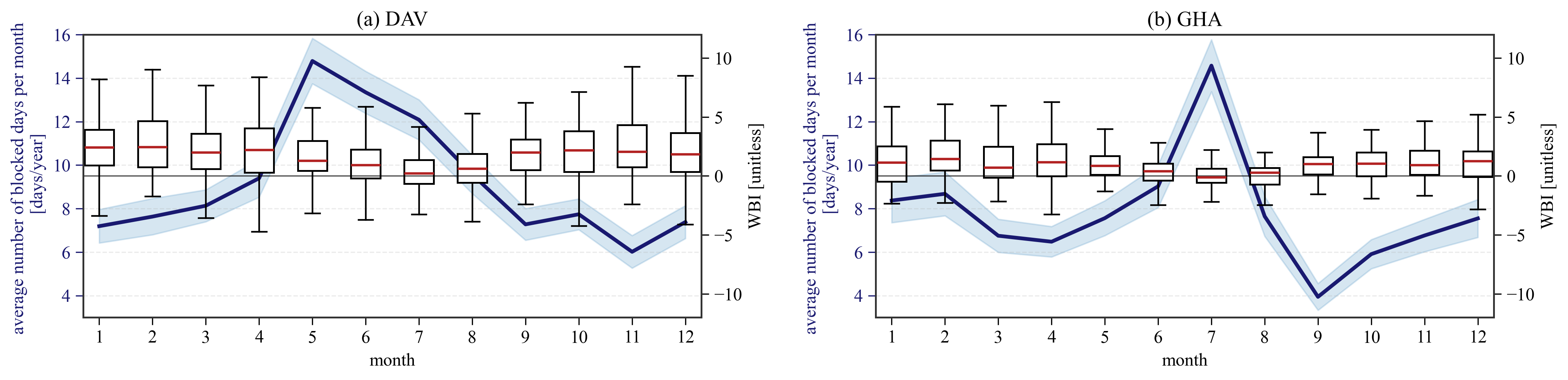}
    \caption{ERA5 annual cycle of WBI compared to blocking frequency expressed as number of blocked days per month. Panel (a) refers to the analysis conducted through the DAV index, while panel (b) refers to the GHA index. In each panel, the blue line expresses the number of blocked days per month averaged over the 1940--2024 reanalysis dataset [days/year]. The blue shadings represent the interannual standard deviation of the number of blocked days. The boxplots showcase the WBI [unitless] distribution for each month.}
    \label{WBI_annual_cycle}
\end{figure}

Figure~\ref{WBI_annual_cycle} compares the annual cycle of blocking frequency—expressed as the number of blocked days per month—with the monthly distribution of WBI.

Both the DAV and GHA indices exhibit a peak of atmospheric blocking frequency during the summer months, with the DAV index capturing this peak slightly earlier than the GHA index. The number of blocked days per month is comparable between the two indices, suggesting that the detection thresholds are well calibrated.

For both indices, the WBI boxplots shift toward negative values during summer. When using the GHA index, the WBI distribution also becomes narrower in these months, indicating reduced variability. A similar, albeit less pronounced, narrowing is observed with the DAV index.

From a dynamical perspective, the shift of the WBI toward negative values in summer reflects an increase in anticyclonic wave breaking (i.e., upstream blocking trajectories), which accounts for nearly half of all events in July. In contrast, during winter, most blocking events are retrograding. As discussed in the Conclusions, this seasonal behavior of the WBI distribution, as well as the varying number of monthly blocked days, may be linked to the latitudinal migration of the polar jet stream, which shifts poleward in summer and equatorward in winter, interacting differently with surface orography and surface thermal gradients.

\subsection{CMIP6 models}     
\label{appendixB}

Table~\ref{tab:models_table} reports the list of climate models that are part of the Coupled Model Intercomparison Project Phase 6 and that we employ in our analysis. We chose those models for which  daily data for both the historical and projections runs were available. For each model, we use the main physics configuration (p1), while incorporating  some members with varying forcing configurations (f2,f3) to increase ensemble size. Detailed forcing specifications  are  not described in more detail here as model dependent. Comparing the number of historical and SSP3-7.0 runs it is evident how the number of runs analyzed varies considerably between models and investigated period. For most of the models, we analyzed more historical runs than projections runs in order to have a good estimate of Greenland blocking variability. Moreover, the model for which we have the largest number of runs is UKESM1-0-LL. Since the results have a greater dependence on the models that provide a larger amount of simulations, we evaluated the impact of excluding UKESM1-0-LL from our analysis. No appreciable differences were found, meaning that UKESM1-0-LL behaves similarly to the average of the other models.

\begin{table}
\centering
\caption{Number of historical and SSP3-7.0 runs per model}
\begin{tabular}{l r r}
\toprule
Model & Historical runs & SSP3-7.0 runs \\
\midrule
ACCESS-CM2      &  3 & 1  \\
CNRM-ESM2-1    &  3 & 3  \\
CanESM5         &  8 & 2  \\
IITM-ESM        &  1 & 1  \\
INM-CM5-0       &  9 & 1  \\
KACE-1-0-G      &  3 & 1  \\
MIROC6          &  8 & 1  \\
MPI-ESM1-2-LR  &  8 & 1  \\
MRI-ESM2-0     &  9 & 1  \\
NorESM2-LM     &  2 & 1  \\
NorESM2-MM     &  2 & 1  \\
UKESM1-0-LL    & 13 & 14 \\
\bottomrule
\end{tabular}
\label{tab:models_table}
\end{table}
\noappendix       

\authorcontribution{MF designed and conducted the research under the supervision of SB. MF developed the tracking algorithm and performed most of the computational analyses. JM provided the platform and datasets for the CMIP6 historical and future projection analyses and computed the blocking indices for the models ensemble. MF wrote the manuscript, and both SB and JM contributed to its revision and editing.} 

\competinginterests{The authors declare that they have no competing interests.} 

\dataavailability{The climate model data are all available from the World Climate Research Programme website (WCRP, 2023) (\url{https://esgf-index1.ceda.ac.uk/projects/cmip6-ceda/}) and ERA5 from the Copernicus climate date store (Climate Data Store, 2023) (\url{https://cds.climate.copernicus.eu/cdsapp\#!/dataset/reanalysis-era5-complete}). The tracking algorithm for the detection of atmospheric blocking adopted in this study is available on GitHub at \url{https://github.com/michele-filippucci/blocktrack} (last access: 3 July 2024; \url{https://doi.org/10.5281/zenodo.13837897}, Filippucci, 2024).}


\begin{acknowledgements}
This research was conducted by MF within the Italian national inter-university doctoral program \emph{Sustainable Development and Climate Change} (\href{https://www.phd-sdc.it/}{https://www.phd-sdc.it}). SB acknowledges support from the National Recovery and Resilience Plan (NRRP), Mission 4, Component 2, Investment 1.4 (Call for Tender No.~1031 of 17/06/2022) of the Italian Ministry for University and Research, funded by the European Union--NextGenerationEU (Project No.~CN\_00000013) and the Italian Ministry of University and Research in the framework of the PRIN 2022 call - Protocol no. 2022WT939B, CUP E53C24002980006. This work made use of JASMIN, the UK’s collaborative data analysis environment (\href{https://www.jasmin.ac.uk/}{https://www.jasmin.ac.uk}) \citep{lawrence2013storing}.
\end{acknowledgements}3

\bibliographystyle{copernicus}
\bibliography{export.bib}

\end{document}